\documentclass[10pt]{article}

\usepackage{graphicx}
\usepackage{graphics}
\usepackage{amscd}
\usepackage{amsmath}
\usepackage{amsfonts}
\usepackage{amssymb}
\usepackage{cite}

\DeclareMathOperator{\wt}{wt_H}

\newcommand{\bF}{ {\mathbb F}}

\newcommand{\C}{ {\mathcal C}}

\newtheorem{theorem}{Theorem}[section]

\newtheorem{corollary}[theorem]{Corollary}

\newtheorem{example}[theorem]{Example}

\newtheorem{lemma}[theorem]{Lemma}

\newtheorem{proposition}[theorem]{Proposition}
\newtheorem{remark}[theorem]{Remark}

\begin{document}

\title{ Binary linear codes with few weights from Boolean functions
\footnote{*Corresponding author.\,\, E-Mail addresses:
 waxiqq@163.com (X.\ Wang), dzheng@hubu.edu.cn(D.\ Zheng), Zhangyan@hubu.edu.cn(Y.\ Zhang)}}

\author{ Xiaoqiang Wang$^1$, Dabin Zheng*$^1$ and Yan Zhang$^2$}

\date{\small 1. Hubei Key Laboratory of Applied Mathematics, \\
Faculty of Mathematics and Statistics, Hubei University, Wuhan 430062, China \\
2. School of Computer Science and Information Engineering, Hubei University, Wuhan 430062, China
}

\maketitle

\leftskip 1.0in
\rightskip 1.0in

\noindent {\bf Abstract.}
Boolean functions have very nice applications in cryptography and coding theory, which have led to a lot of research focusing on their applications. The objective of this paper is to construct binary linear codes with few weights from the defining set, which is defined by some special Boolean functions and some additional restrictions.
First, we provide two general constructions of binary linear codes with three or four weights from Boolean functions with at most three Walsh transform values and determine
the parameters of their dual codes. Then many classes of binary linear codes with explicit weight enumerators are obtained. Some binary linear codes and their duals obtained are optimal or almost optimal. The binary linear codes obtained in this paper may have a special interest in secret sharing schemes, association schemes, strongly regular graphs.



\vskip 35pt

\leftskip 0.0in
\rightskip 0.0in

\section{ Introduction}

Let $q$ be a prime power and $n$ be a positive integer. An $[n, k, d]$ code $\mathcal{C}$ over the finite field $\mathbb{F}_q$ is a $k$-dimensional linear subspace of $\mathbb{F}_q^n$ with minimum Hamming distance $d$. Let $A_i$ denote the number of codewords with Hamming weight $i$ in $\C$. The {\it weight enumerator} of $\C$ is defined by
$1+ A_1 x+ A_2x^2 + \cdots + A_nx^n$ and the sequence $(1, A_1, A_2, \cdots, A_n)$ is called the {\it weight distribution} of $\C$. A code $\C$ is said to be a $t$-weight code if the number of nonzero $A_i$ in the sequence $(1, A_1, A_2, \cdots, A_n)$ is equal to $t$. An $[n, k, d]$ code over $\mathbb{F}_q$ is called {\it distance-optimal} if there is no $[n,k,d+1]$ code over  $\mathbb{F}_q$, and {\it dimension-optimal} if there is no $[n,k+1,d]$ code over  $\mathbb{F}_q$. A code is said to be optimal if it is both distance-optimal and dimension-optimal.

Binary error correcting linear codes are widely studied by researchers and employed by engineers since they have applications in computer and communication systems, data storage devices and consumer electronics. In particular, due to linear codes with few weights have applications in secret sharing~\cite{A1998,Carlet2005,Cohen2016,Yuan2006}, strongly regular graphs~\cite{CA1986}, association schemes~\cite{CA1984} and authentication codes~\cite{Ding2005}, many researchers focused on constructions of linear codes with few weights and made a lot of progress on this topic. A non-exhaustive list dealing with linear codes with few weights
is~\cite{DingNieder2007, Ding2015, Ding2016,DDing2014,HengYue2015,HengWang2020,Jian2019,LiYueFu2016,Li2020,
LuoCaoetal2018,Mesnager2017,TangLietal2016,Wangetal2015,Wang2015,Wangetal2016,Wu2020,Xiaetal2017,ZhengBao2017,ZhouLietal2015}. Almost  all known linear codes in the the previous literature were constructed by trace representations. As far as we know, Ding et al.~\cite{DingNieder2007} first constructed a generic class of linear codes
by trace representations as follows:
\begin{equation*}
\C_D = \left\{ \left( {\rm Tr}_1^m( xd_1), {\rm Tr}_1^m( xd_2), \cdots, {\rm Tr}_1^m(xd_n) \right) \, |\, x\in \bF_{p^m} \right\},
\end{equation*}
where ${\rm Tr}_1^m$ denote the trace function from $\bF_{p^m}$ to $\bF_p$ and $D=\{d_1, d_2, \ldots, d_n\}\subset \bF_{p^m}$. The code $\C_D$ is a linear code over $\bF_p$ with dimension at most $m$ and $D$ is called the defining set of $\C_D$. Along this line, Li et al.\cite{LiYueFu2016} considered a class of linear codes with dimension
at most $2m$ of the form
\begin{equation}\label{eq:defc}
\begin{split}
\mathcal{C}_{D}=\left\{c(a,b)=\left({\rm Tr}_1^m(ax+by)\right)_{(x ,y)\in {D}}: \, a,b \in \mathbb{F}_{p^m} \right\}
\end{split}
\end{equation}
and studied $\C_D$ for the case $D= \left\{ (x, y)\in \bF_{p^m}^2\setminus \{ (0,0)\} \, :\, {\rm Tr}_1^m\left(x^{N_1} + y^{N_2}\right)= 0 \right\}$, where $N_1, N_2 \in \left\{ 1,\,\,2,\,\, p^{\frac{m}{2}+1}\right\}$.
Then, this construction was generalized to the other cases of $D$ by Jian et al.~\cite{Jian2019} and Li~\cite{Li2020}, and some  linear codes with few weights were obtained.
Very recently,  Wu et al.~\cite{Wu2020} studied the $p$-ary linear code $\C_D$ for the case $D=\{(x,y)\in \mathbb{F}_{p^m}^2\setminus\{(0,0)\}: \, f(x)+g(y)=0\}$ for any odd prime $p$, where $f(x)={\rm Tr}_1^m(x)$ and $g(y)$ is a weakly regular bent function, or both $f(x)$ and $g(y)$ are weakly regular bent functions.

Inspired by the works in~\cite{Wu2020}, this paper considers binary linear codes of the form (\ref{eq:defc}) by employing some special Boolean functions and
more restrictions on defining sets. Concretely, we first study the linear codes of the form (\ref{eq:defc}) by selecting the defining set as
\begin{equation}\label{eq:defc1}
\begin{split}
D_{\epsilon}=\left\{ (x,y)\in \mathbb{F}_{2^m}^2\setminus\{(0,0)\}: \, f(x)+g(y)=0\,\, \text{and}\,\,{\rm Tr}_1^m(x+y)=\epsilon \right\},
\end{split}
\end{equation}
where $\epsilon \in \{ 0, 1\}$ and $f(x)$ and $g(y)$ are Boolean functions from $\bF_{2^m}$ to $\bF_2$ with at most three Walsh transform values. We call the linear codes obtained from the definition set (\ref{eq:defc1}) the first class of linear codes. When the Walsh spectra of $f(x)$ and $(y)$ satisfy some conditions, we determine the weight
distribution of $\C_{D_{ \epsilon}}$ and the parameters of their dual codes for $ \epsilon \in \left\{ 0, 1\right\}$. The second contribution of this paper is that we derive new at most three or four weight linear codes of the form~(\ref{eq:defc}) from the following defining set:
\begin{equation}\label{eq:defc2}
\begin{split}
D_{ \epsilon}=\left\{(x,y)\in \mathbb{F}_{2^m}^2\setminus\{(0,0)\}: \, f(x)+g(y)=0,\,\, {\rm Tr}_1^m(x)=0\,\,\text{and}\,\,{\rm Tr}_1^m(y)=\epsilon\right\},
\end{split}
\end{equation}
where $f(x)$ and $g(y)$ are Boolean functions with at most three Walsh transform values satisfying some additional conditions. We call the linear codes obtained from the definition set (\ref{eq:defc2}) the second class of linear codes. Some of  binary linear codes obtained in this paper are optimal or almost optimal.

The rest of this paper is organized as follows. In Section~2, we introduce some preliminaries. Section~3 introduces the Walsh transform values of some quadratic Boolean functions. In Section~4, we investigate the weight distribution of the first class of linear codes and the parameters of their dual codes. Section~5 investigates the weight distribution of the second class of linear codes and the parameters of their dual codes. Section~6 concludes this paper.

\section{ Preliminaries}

Throughout this paper, we adopt the following notation unless otherwise stated:
\begin{description}
\item{$\bullet$} $\bF_{2^m}$ is a finite field with $2^m$ elements.
\item{$\bullet$} ${\rm Tr}_{\ell}^m(\cdot)$ is the trace function from $\bF_{2^m}$ to $\bF_{2^\ell}$,  where $\ell, m$ are positive integers with $\ell\, | \, m$.
\item{$\bullet$} $v_2(\cdot)$ is the 2-adic order function and set $v_2(0)=\infty$.
\item{$\bullet$} ${\rm T}_k^{\ell k}(x):=\sum_{i=0}^{\ell-1}x^{2^{ik}}$, where $x$ is a variable.
\item{$\bullet$} ${\rm T}_u^v\circ {\rm T}_{u_0}^{v_0}(x)={\rm T}_u^v( {\rm T}_{u_0}^{v_0}(x))$,  where $u,v, u_0$ and $v_0$ are positive integers with $u\,|\,v$ and $u_0\,|\,v_0$.
\end{description}

\begin{lemma}[\cite{Mesnger2020}]\label{lemtr}
 Follow the notation introduced above. Denote $d=\gcd(\ell k,m)$ and let $a \in \mathbb{F}_{2^m}$. The equation ${\rm T}_k^{ \ell k}(x)=a$ has a solution in $\mathbb{F}_{2^m}$ if and only if   ${\rm T}_1^{(d,k)}\circ {\rm T}_1^{2} \circ {\rm T}_d^m(a)=0$ when $\frac{\ell k}{[d,k]}$ is odd and ${\rm T}_d^{m}(a)=0$ when $\frac{\ell k}{[d,k]}$ is even, where $[d,k]$ is the lowest common multiple of two positive
integers $d$ and $k$.
\end{lemma}

 For convenience, we  introduce a few basic concepts, which will be used in the following sections. Let $f(x)$ be a Boolean function from $\mathbb{F}_{2^m}$ to $\mathbb{F}_2$. The {\it Walsh transform} of $f(x)$ is defined by
\begin{equation}\label{eq:walshtransform}
\hat{f}(\omega)=\sum_{x \in \mathbb{F}_{2^m}}(-1)^{f(x)+{\rm Tr}_1^m(\omega x)},\,\, \omega \in \mathbb{F}_{2^m}.
\end{equation}
\begin{description}
\item{$\bullet$} If $f(x)$ satisfies $\hat{f}(\omega)\in\{\pm 2^{\frac{m}{2}}\}$ for all $w \in \mathbb{F}_{2^m}$, then $f(x)$ is called a {\it bent function}.  Bent functions were coined by Rothaus in \cite{Rothaus1976} and exist only for even $m$.
\item{$\bullet$} If $m$ is odd and $f(x)$ satisfies $\hat{f}(\omega)\in \{0, \pm2^{\frac{m+1}{2}}\}$ or  $m$ is even and $f(x)$ satisfies $\hat{f}(\omega)\in \{0, \pm2^{\frac{m+2}{2}}\}$ for all $w \in \mathbb{F}_{2^m}$, then $f(x)$ is called a {\it semibent function} \cite{Mesnager2011}.
\item{$\bullet$} If $f(x)$ satisfies $\hat{f}(\omega)\in \{0, \pm A\}$  for all $w \in \mathbb{F}_{2^m}$, then $f(x)$ is called a {\it plateaued function}. By {\it Parseval's identity}, then $A= 2^{\frac{m+d}{2}}$, where $d$ is an integer such that $0\leq d \leq m$. Clearly, bent functions and almost bent functions are the special cases of plateaued functions \cite{Cesmelioglu2013}.
\end{description}




 To study the parameters of the dual codes of the objective linear codes, we need the Pless power moment
identities on linear codes. Let $\mathcal{C}$ be a binary $[n, k]$ code, and denote its dual by $\mathcal{C}^{\perp}$. Let
 $A_i$ and $A^{\perp}_i$ be the number of codewords of weight $i$ in $\mathcal{C}$ and $\mathcal{C}^{\perp}$, respectively. Then we have the first four Pless power moments identities (\cite{MacWilliam1997}, p. 131) as follows:
\begin{equation*}
\begin{split}
&\sum_{i=0}^nA_i=2^k;\,\,\, \sum_{i=0}^niA_i=2^{k-1}(n-A_1^{\perp});\,\,\,\sum_{i=0}^ni^2A_i=2^{k-2}[n(n+1)-2nA_1^{\perp}+2A_2^{\perp}];\\
&\sum_{i=0}^ni^3A_i=2^{k-3}[n^2(n+3)-(3n^2+3n-2)A_1^{\perp}+6nA_2^{\perp}-6A_3^{\perp}].\\
 \end{split}
\end{equation*}

The following is a well-known result.


\begin{lemma}[Sphere Packing Bound]
Let $\mathcal{C}$ be a binary $[n, k,d]$ code. Then
$$ 2^n\geq 2^k\sum_{i=0}^{\lfloor \frac{d-1}{2} \rfloor}\left(
\begin{array}{cccc}
   n  \\
     i  \\
\end{array}
\right),$$
where $\lfloor \frac{d-1}{2} \rfloor$ is the largest integer less than or equal to $\frac{d-1}{2}$.
\end{lemma}

\section{ Walsh transform values of some quadratic Boolean functions }

Let $f(x)$ be a Boolean function from $\mathbb{F}_{2^m}$ to $\mathbb{F}_2$ and its Walsh transform defined in (\ref{eq:walshtransform}). The Walsh transform was used to characterize some properties of Boolean functions, such as nonlinearity, balance, etc.. Boolean functions with few Walsh transform values were extensively studied due to their applications in cryptography, error correcting codes and signal sequence design. However, as far as we know, there is few research on study of the relation between two Walsh transform values of Boolean functions. The following lemmas show that there exist quadratic Boolean functions $f(x)$ such that $\hat{f}(\omega)\hat{f}(\omega+1)=0$ for any $\omega\in\mathbb{F}_{2^m}$.

\begin{lemma}\label{beglem12}
Let $m, k$ be positive integers with $d=\gcd(m,k)$ and $v_2(\cdot)$ denote the 2-adic order function. Let $f(x)={\rm Tr}_1^m(\alpha x^{2^k+1})$ be a Boolean function from $\mathbb{F}_{2^m}$ to $\mathbb{F}_2$ for some $\alpha \in \mathbb{F}_{2^m}^*$. If $\alpha \in \{ c^{2^k+1}\, | \, c \in \mathbb{F}_{2^m}^*\}$, i.e., there exists $\beta \in \mathbb{F}_{2^m}^*$ such that $\alpha=\beta^{2^k+1}$, then
\begin{equation*}
\begin{split}
\hat{f}(\omega)=\begin{cases}
\pm 2^{\frac{m+d}{2}}& \text{if \,\,$v_2(m)\leq v_2(k)$ and   ${\rm Tr}_d^m(\omega\beta^{-1})=1$, } \\
\pm 2^{\frac{m+2d}{2}}& \text{if \,\,$v_2(m)\geq v_2(k)+1$ and ${\rm Tr}_{2d}^m(\omega\beta^{-1})=0$, } \\
0& \text{otherwise.}
\end{cases}
\end{split}
\end{equation*}
When $v_2(m)\leq v_2(k)$, $\hat{f}(\omega)\hat{f}(\omega+1)=0$ for any $\omega\in\mathbb{F}_{2^m}$ if and only if ${\rm Tr}_d^m(\beta^{-1})\neq 0$.
When $v_2(m)\geq v_2(k)+1$, $\hat{f}(\omega)\hat{f}(\omega+1)=0$ for any $\omega\in\mathbb{F}_{2^m}$ if and only if ${\rm Tr}_{2d}^m(\beta^{-1})\neq 0$.
\end{lemma}
{\it Proof.} Note that $\{x \in \mathbb{F}_{2^m} \, | \, x^{2^d+1}\}=\{x \in \mathbb{F}_{2^m} \, | \, x^{2^k+1}\}$ since $d=\gcd(k,m)$, then the possible values of $\hat{f}(\omega)$ can be easily obtained from \cite{Coulter1999, Coulter2002}. Now we consider the necessary and sufficient condition of $\hat{f}(\omega)\hat{f}(\omega+1)=0$ for any $\omega\in \mathbb{F}_{2^m}$.

When $v_2(m)\leq v_2(k)$, it is obvious there are some $\omega$ such that ${\rm Tr}_d^m(\omega\beta^{-1})=1$ for  $\beta \in \mathbb{F}_{2^m}^*$. Then $\hat{f}(\omega)\hat{f}(\omega+1)=0$ for any $\omega\in \mathbb{F}_{2^m}$ if and only if one of ${\rm Tr}_d^m(\omega\beta^{-1})$ and ${\rm Tr}_d^m((\omega+1)\beta^{-1})$ is equal to $1$, i.e., ${\rm Tr}_d^m(\beta^{-1})\neq0$.
When $v_2(m)\geq v_2(k)+1$, the results can be shown similarly. $\square$

\begin{remark}
If $\alpha \notin  \{ c^{2^k+1}\, |\, c \in \mathbb{F}_{2^m}^*\}$, then $f(x)={\rm Tr}_1^m(\alpha x^{2^k+1})$ is a Gold bent function and $\hat{f}(\omega)=  \pm 2^{\frac{m}{2}}$ for any $\omega \in \mathbb{F}_{2^m}$.
\end{remark}

\begin{lemma}\label{begle}
Let $f(x)={\rm Tr}_1^m(\sum_{i=1}^{\ell}x^{2^{ik}+1})$  be a Boolean function  from $\mathbb{F}_{2^m}$ to $\mathbb{F}_2$. Assume that  $\gcd(\ell k,m)=\gcd((\ell+1)k,m)=1$, then $\hat{f}(\omega)\cdot \hat{f}(\omega+1)=0$ for any $\omega \in \mathbb{F}_{2^m}$, and the possible values of $\hat{f}(\omega)$ are given as follows:
\begin{equation*}
\begin{split}
\hat{f}(\omega)=\begin{cases}
\pm 2^{\frac{m+1}{2}},& \text{if \,\,${\rm Tr}_1^m(\ell+\omega)=0$,} \\
0,& \text{if \,\,${\rm Tr}_1^m(\ell+\omega)=1$.} \\
\end{cases}
\end{split}
\end{equation*}
\end{lemma}
{\it Proof.} It is clear that
\begin{equation}\label{eq:31214}
\begin{split}
\hat{f}^2(\omega)&=\sum_{x_0 \in \mathbb{F}_{2^m}}(-1)^{{\rm Tr_1^m}\left(\sum_{i=1}^{\ell}x_0^{2^{ik}+1}+\omega x_0\right)}\sum_{x \in \mathbb{F}_{2^m}}(-1)^{{\rm Tr}_1^m\left(\sum_{i=1}^\ell x^{2^{ik}+1}+\omega x\right)}\\
&=\sum_{x, y \in \mathbb{F}_{2^m}}(-1)^{{\rm Tr}_1^m\left(\sum_{i=1}^\ell(x+y)^{2^{ik}+1}+\omega (x+y)+\sum_{i=1}^\ell x^{2^{ik}+1}+\omega x\right)}\\
&=\sum_{x, y \in \mathbb{F}_{2^m}}(-1)^{{\rm Tr}_1^m\left(\sum_{i=1}^\ell xy^{2^{ik}}+\sum_{i=1}^lx^{2^{ik}}y+\sum_{i=1}^\ell y^{2^{ik}+1}+\omega y\right)}\\
&=\sum_{y \in \mathbb{F}_{2^m}}(-1)^{{\rm Tr}_1^m\left(\sum_{i=1}^\ell y^{2^{ik}+1}+\omega y\right)}\sum_{x \in  \mathbb{F}_{2^m}}(-1)^{{\rm Tr}_1^m\left(\sum_{i=1}^\ell xy^{2^{ik}}+\sum_{i=1}^lx^{2^{ik}}y\right)}\\
&=\sum_{y \in \mathbb{F}_{2^m}}(-1)^{{\rm Tr}_1^m\left(\sum_{i=1}^\ell y^{2^{ik}+1}+\omega y\right)}\sum_{x \in  \mathbb{F}_{2^m}}(-1)^{{\rm Tr}_1^m\big(\big(z+z^{2^{(\ell+1)k}}\big)x^{2^{\ell k}}\big)}\\
&=2^m\sum_{y \in \mathbb{F}_{2^m}, \,\,z+z^{2^{(\ell+1)k}}=0}(-1)^{{\rm Tr}_1^m\left(\sum_{i=1}^\ell y^{2^{ik}+1}+\omega y\right)},\\
\end{split}
\end{equation}
where $z=y+y^{2^k}+\cdots+y^{2^{(\ell-1)k}}$.
It is easy to see that $z+z^{2^{(\ell+1)k}}=0$ if and only if $z=0$ or $z=1$ since $\gcd((\ell+1)k,m)=1$. Hence, we have
\begin{equation*}
\hat{f}^2(\omega)=2^m\sum_{y\in \mathbb{F}_{2^m}, z\in \mathbb{F}_2}(-1)^{{\rm Tr}_1^m\left(\sum_{i=1}^\ell y^{2^{ik}+1}+\omega y\right)}.
\end{equation*}
Next, we discuss the values of $\hat{f}^2(\omega)$ for $\omega$ running through $\bF_{2^m}$.

\noindent{\bf Case 1:}  $\ell$ is odd. As ${\rm T}_1^2\circ {\rm T}_1^m(x)=0$ for any $x \in \mathbb{F}_{2^m}$, by Lemma \ref{lemtr}, $y+y^{2^k}+\cdots+y^{2^{(\ell-1)k}}=a$ has solutions for all $a\in \mathbb{F}_{2^m}$. It is obvious that for different elements $a_0,a_1\in \mathbb{F}_{2^m}$, the solutions
    $y+y^{2^k}+\cdots+y^{2^{(\ell-1)k}}=a_0$ and  $y+y^{2^k}+\cdots+y^{2^{(\ell-1)k}}=a_1$ are different. Hence,
    $y+y^{2^k}+\cdots+y^{2^{(\ell-1)k}}=0$ and $y+y^{2^k}+\cdots+y^{2^{(\ell-1)k}}=1$ have only one solution, respectively. Clearly, $y=0$ or $y=1$ is the solution of $y+y^{2^k}+\cdots+y^{2^{(\ell-1)k}}=0$ or $y+y^{2^k}+\cdots+y^{2^{(\ell-1)k}}=1$, respectively. Hence,
     \begin{equation*}
\begin{split}
\hat{f}^2(\omega)=2^m\sum_{y\in \mathbb{F}_2}(-1)^{{\rm Tr}_1^m\left(\sum_{i=1}^\ell y^{2^{ik}+1}+\omega y\right)}=2^m\left(1+(-1)^{{\rm Tr}_1^m(\ell+\omega)}\right).
\end{split}
\end{equation*}

\noindent{\bf Case 2:}  $\ell$ is even. As $\gcd(\ell k,m)=1$, then $m$ must be odd. By Lemma \ref{lemtr}, $y+y^{2^k}+\cdots+y^{2^{(\ell-1)k}}=a$ has solutions if and only if ${\rm T}_1^m(a)=0$. As $y+y^{2^k}+\cdots+y^{2^{(\ell-1)k}}$ is a linear polynomial and the number of $a\in \mathbb{F}_{2^m}$ such that ${\rm T}_1^m(a)=0$ is $2^{m-1}$, then the equation $y+y^{2^k}+\cdots+y^{2^{k(\ell-1)}}=0$ has two solutions and $y+y^{2^k}+\cdots+y^{2^{(\ell-1)k}}=1$ has not solution. Clearly, $y=0$ and $y=1$ are the solutions of $y+y^{2^k}+\cdots+y^{2^{(\ell-1)k}}=0$. Hence,
     \begin{equation*}
\begin{split}
\hat{f}^2(\omega)=2^m\sum_{y\in \mathbb{F}_2}(-1)^{{\rm Tr}_1^m\left(\sum_{i=1}^\ell y^{2^{ik}+1}+\omega y\right)}=2^m\left(1+(-1)^{{\rm Tr}_1^m(\ell+\omega)}\right).
\end{split}
\end{equation*}
Therefore, no matter $\ell$ is odd or even, we have $\hat{f}^2(\omega)=2^m(1+(-1)^{{\rm Tr}_1^m(\ell+\omega)})$. As $\gcd(\ell k,m)=\gcd((\ell+1)k,m)=1$, then $m$ is odd and $\hat{f}(\omega)\cdot \hat{f}(\omega+1)=0$ for any $\omega \in \mathbb{F}_{2^m}$, the desired conclusion follows.  $\square$

In fact, if we do not put such strong restrictions on the Boolean function $f(x)$ in Lemma \ref{begle}, the Walsh transform values of $f(x)$ still satisfy  $\hat{f}(\omega)\cdot \hat{f}(\omega+1)=0$ for any $\omega \in \mathbb{F}_{2^m}$.
\begin{lemma}\label{begle1}
Let $v_2(\cdot)$ denote the 2-adic order function and $f(x)={\rm Tr}_1^m(\sum_{i=1}^{\ell}x^{2^{ik}+1})$  be a Boolean function from $\mathbb{F}_{2^m}$ to $\mathbb{F}_2$. If $v_2(m)\leq v_2((\ell+1)k)$, then $\hat{f}(\omega)\cdot \hat{f}(\omega+1)=0$ for any $\omega \in \mathbb{F}_{2^m}$.
\end{lemma}
{\it Proof.} By similar computations as in (\ref{eq:31214}),  we  obtain
\begin{equation*}
\hat{f}(\omega)\hat{f}(\omega+1)=2^m\sum_{y \in \mathbb{F}_{2^m}, \,\,z+z^{2^{(\ell+1)k}}=1}(-1)^{{\rm Tr}_1^m\left(\sum_{i=1}^\ell y^{2^{ik}+1}+\omega y\right)},
\end{equation*}
where $z=y+y^{2^k}+\cdots+y^{2^{(\ell-1)k}}$. Assume that $d=\gcd((\ell+1)k,m)$, then ${\rm Tr}_d^m(z+z^{2^{(\ell+1)k}})=0$, which is contradict to $z+z^{2^{(\ell+1)k}}=1$ since $v_2(m)\leq v_2((\ell+1)k)$. This means that there does not exist $y\in \mathbb{F}_{2^m}$ such that $z+z^{2^{(\ell+1)k}}=1$. Hence, $\hat{f}(\omega)\hat{f}(\omega+1)=0$.

\begin{remark}
From Lemmas~\ref{beglem12}, \ref{begle} and~\ref{begle1}, we see that there exist some Boolean functions $f(x)$ such that $\hat{f}(\omega)\hat{f}(\omega+1)=0$ for any $\omega \in \mathbb{F}_{2^m}$. Such Boolean functions will be used to construct binary linear codes with few weights in Section~$4$ and Section $5$.
\end{remark}

\section{The weight distribution of the first class of linear codes}

In this section, we investigate the weight distribution of the linear code $\mathcal{C}_{D_{ \epsilon}}$, where $\mathcal{C}_{D_{ \epsilon}}$ has the form (\ref{eq:defc}) and $D_{ \epsilon}$ is defined in (\ref{eq:defc1}). Assume that $n=|D_{ \epsilon}|$ is the length of $\C_{D_{ \epsilon}}$, then
\begin{equation}\label{lengthn}
\begin{split}
n&=\sum_{(x,y)\in \mathbb{F}_{2^m}^2\setminus\{(0,0)\}} \left({\frac{1}{2}}\sum_{z_0 \in \mathbb{F}_{2}}(-1)^{z_0\left(f(x)+g(y)\right)}\right)\left({\frac{1}{2}}\sum_{z_1 \in \mathbb{F}_{2}}(-1)^{z_1\left({\rm Tr}_1^m(x+y)-\epsilon\right)}\right)\\
&=\sum_{(x,y)\in \mathbb{F}_{2^m}^2}\left({\frac{1}{2}}\sum_{z_0 \in \mathbb{F}_{2}}(-1)^{z_0\left(f(x)+g(y)\right)}\right)\left({\frac{1}{2}}\sum_{z_1 \in \mathbb{F}_{2}}(-1)^{z_1({\rm Tr}_1^m(x+y)-\epsilon)}\right)-\delta\\
&=\frac{1}{4}\sum_{(x,y)\in \mathbb{F}_{2^m}^2} \left( (-1)^{f(x)+g(y)}+1\right)\left((-1)^{{\rm Tr}_1^m(x+y)-\epsilon}+1\right)-\delta\\
&=2^{2m-2}+\frac{1}{4}\sum_{x \in \mathbb{F}_{2^m}}\sum_{y \in \mathbb{F}_{2^m}}(-1)^{f(x)+g(y)}+\frac{1}{4}(-1)^{\epsilon}\sum_{x \in \mathbb{F}_{2^m}}\sum_{y \in \mathbb{F}_{2^m}}(-1)^{{\rm Tr}_1^m(x+y)}\\
&+\frac{1}{4}(-1)^{\epsilon}\sum_{x \in \mathbb{F}_{2^m}}(-1)^{f(x)+{\rm Tr}_1^m(x)}\sum_{y \in \mathbb{F}_{2^m}}(-1)^{g(y)+{\rm Tr}_1^m(y)}-\delta\\
&=2^{2m-2}+\frac{1}{4}\hat{f}(0)\hat{g}(0)+\frac{1}{4}(-1)^{\epsilon}\hat{f}(1)\hat{g}(1)-\delta,
\end{split}
\end{equation}
where
\begin{equation}\label{eq:11t}
\delta=\frac{1}{2}\sum_{z_1 \in \mathbb{F}_2}(-1)^{z_1\epsilon}
=\left\{ \begin{array}{lll}
1, \,\,\text{if $\epsilon=0$},\\
0, \,\,\text{if $\epsilon=1$.}
\end{array}\right.
\end{equation}
For any  $(a,b)\in \mathbb{F}_{2^m}^2$, the Hamming weight of the codeword $\mathbf{c}(a,b)=\left( {\rm Tr}_1^m(ax+by)\right)_{(x,y)\in D_{ \epsilon}}$ in $\mathcal{C}_{D_{ \epsilon}}$ is
\begin{equation}\label{eq:wt0}
{\rm wt_H}( {\bf c}(a,b))=n-N(a,b),
\end{equation}
where $n$ is the length of the linear code $\mathcal{C}_{D_{\epsilon}}$ and
\begin{equation*}
\begin{split}
N(a,b)&=\left|\left\{(x,y)\in \mathbb{F}_{2^m}^2\setminus\{(0,0)\}:\,\, f(x)+g(y)=0, \,\, {\rm Tr}_1^m\left(x+y\right)=\epsilon \,\,\text{and}\,\, {\rm Tr}_1^m\left(ax+by\right)=0 \right\}\right|. \\
\end{split}
\end{equation*}
 From the definition of $n$, it is easy to see that when $(a,b)=(1,1)$, we have
\begin{equation}\label{eq:wt01}
N(a,b)=\left\{ \begin{array}{lcl}
n, \,\,\text{if $\epsilon=0$},\\
0, \,\,\text{if $\epsilon=1$.}
\end{array}\right.
\end{equation}
If $(a,b)\neq(0,0)$ and $(a,b)\neq(1,1)$, then
\begin{equation}\label{eqNab}
\begin{split}&N(a,b)\\
&=\sum_{(x,y)\in \mathbb{F}_{2^m}^2\setminus\{(0,0)\}} \frac{1}{2}\left(\sum_{z_0 \in \mathbb{F}_{2}}(-1)^{z_0(f(x)+g(y))}\right)\frac{1}{2}\left(\sum_{z_1 \in \mathbb{F}_{2}}(-1)^{z_1({\rm Tr}_1^m(x+y)-\epsilon )}\right)\frac{1}{2}\left(\sum_{z_2 \in \mathbb{F}_{2}}(-1)^{z_2\left({\rm Tr}_1^m(ax+by)\right)}\right)\\
&=\frac{1}{8}\sum_{(x,y)\in \mathbb{F}_{2^m}^2} \left(1+(-1)^{f(x)+g(y)}\right)\left(1+(-1)^{{\rm Tr}_1^m(x+y)-\epsilon}\right)\left(1+(-1)^{{\rm Tr}_1^m(ax+by)}\right)-\delta\\
&=2^{2m-3}+\frac{(-1)^{\epsilon}}{8}\sum_{x,y\in \mathbb{F}_{2^m}}(-1)^{{\rm Tr}_1^m(x+y)}+\frac{1}{8}\sum_{x, y \in \mathbb{F}_{2^m}}(-1)^{{\rm Tr}_1^m(ax+by)}+\frac{1}{8}\sum_{x,y\in \mathbb{F}_{2^m}}(-1)^{f(x)+g(y)}\\
&+\frac{(-1)^{\epsilon}}{8}\sum_{x \in \mathbb{F}_{2^m}}(-1)^{f(x)+{\rm Tr}_1^m((a+1)x)}\sum_{y \in \mathbb{F}_{2^m}}(-1)^{g(y)+{\rm Tr}_1^m((b+1)y)}
+\frac{(-1)^{\epsilon}}{8}\sum_{x \in \mathbb{F}_{2^m}}(-1)^{{\rm Tr}_1^m((a+1)x)}\sum_{y \in \mathbb{F}_{2^m}}(-1)^{{\rm Tr}_1^m((b+1)y)}\\
&+\frac{(-1)^{\epsilon}}{8}\sum_{x \in \mathbb{F}_{2^m}}(-1)^{f(x)+{\rm Tr}_1^m(x)}\sum_{y \in \mathbb{F}_{2^m}}(-1)^{g(y)+{\rm Tr}_1^m(y)}+\frac{1}{8}\sum_{x \in \mathbb{F}_{2^m}}(-1)^{f(x)+{\rm Tr}_1^m(ax)}\sum_{y \in \mathbb{F}_{2^m}}(-1)^{g(y)+{\rm Tr}_1^m(by)}-\delta\\
&=2^{2m-3}+\frac{1}{8}(\hat{f}(0)\hat{g}(0)+\hat{f}(a)\hat{g}(b))+
\frac{(-1)^{\epsilon}}{8}(\hat{f}(1)\hat{g}(1)+\hat{f}(a+1)\hat{f}(b+1))-\delta\\
&=\frac{n-\delta}{2}+\frac{\hat{f}(a)\hat{g}(b)}{8}+
\frac{(-1)^{\epsilon}\hat{f}(a+1)\hat{g}(b+1)}{8},
\end{split}
\end{equation}
where $n$ and $\delta$ are defined in (\ref{lengthn}) and (\ref{eq:11t}), respectively.

With the above preparations, we have the following results.

\begin{proposition}\label{pro1}
Follow the notation introduced above. Assume that $(a,b) \in \mathbb{F}_{2^m}^2\setminus \{(0,0),(1,1)\}$. If $\epsilon=0$, then $\mathcal{C}_{D_{ 0}}$ is a binary linear code of length $n$ and its Hamming weights are given by the following multiset
\begin{equation*}
\left\{\frac{n+1}{2}-\frac{\hat{f}(a)\hat{g}(b)}{8}-\frac{\hat{f}(a+1)\hat{g}(b+1)}{8}\right\}\bigcup\left\{0\right\}.
\end{equation*}
If $\epsilon=1$, then $\mathcal{C}_{D_{ 1}}$ is a binary linear code of length $n$ and its Hamming weights are given by the following multiset
\begin{equation*}
\left\{\frac{n}{2}-\frac{\hat{f}(a)\hat{g}(b)}{8}+\frac{\hat{f}(a+1)\hat{g}(b+1)}{8}\right\}\bigcup\left\{0,n\right\}.
\end{equation*}
\end{proposition}

 In the following, we determine the weight distribution of $\mathcal{C}_{D_{ \epsilon}}$ for some special Boolean functions. For convenience, we write
\begin{equation}\label{eq:phi1}
\Phi_{\epsilon}=\frac{1}{4}\left(\hat{f}(0)\hat{g}(0)+(-1)^{\epsilon}\hat{f}(1)\hat{g}(1)\right).
\end{equation}

\begin{theorem}\label{theorem1}
Let $m$ be an integer with $m\geq 3$ and $\mathcal{C}_{D_{ \epsilon}}$ be a linear code with the defining set $D_{ \epsilon}$ given in~(\ref{eq:defc1}). Let $f(x)$ and $g(y)$ in~(\ref{eq:defc1}) satisfy one of the following conditions:
\begin{description}
\item{{\rm (i)}}  $\hat{f}(a) \in \left\{0, \pm2^{\frac{m+d_0}{2}}\right\}$, $\hat{f}(a)\cdot \hat{f}(a+1)=0$  and $\hat{g}(b)\in \left\{0, \pm2^{\frac{m+d_1}{2}}\right\}$ for any $a,b \in \mathbb{F}_{2^m}$;
\item{{\rm (ii)}}  $\hat{f}(a) \in \left\{0, \pm2^{\frac{m+d_0}{2}}\right\}$, $\hat{f}(a)=\pm \hat{f}(a+1)$, $\hat{g}(b)\in \left\{0, \pm2^{\frac{m+d_1}{2}}\right\}$ and $\hat{g}(b)=\pm \hat{g}(b+1)$ for any $a,b \in
\mathbb{F}_{2^m}$.
\end{description}
Denote $t=\frac{d_0+d_1}{2}$ or $\frac{d_0+d_1+2}{2}$ if condition (i) or (ii) holds, respectively. Assume that $\Phi_{\epsilon}\neq 2^{m+t-2}-2^{2m-2}$, then the following statements hold.
\begin{description}
\item{{\rm (1)}} If $\epsilon=0$, then $\mathcal{C}_{D_{ 0}}$ is an $[n, 2m-1]$ code with weight distribution in Table \ref{Table1}, where
    $n=2^{2m-2}+\Phi_{0}-1.$
Its dual code has parameters $[n, n-2m+1, 3]$.
\begin{table}[h]
{\caption{\rm   The weight distribution of $\mathcal{C}_{D_{0}}$ }\label{Table1}
\begin{center}
\begin{tabular}{cccc}\hline
     Weight & Multiplicity \\\hline
  $0$ & $1$  \\
 $\frac{n+1}{2}$ & $2^{4-2t-2m}(n+1)^2+2^{2m-1}-2^{3-2t}\cdot(n+1)-1$ \\
  $\frac{n+1}{2} + 2^{m-3+t}$ & $(n+1)\cdot 2^{1-m-t}-2^{m-t}-(2^{3-2m}\cdot(n+1)^2-4\cdot(n+1))\cdot2^{-2t}$ \\
    $\frac{n+1}{2} - 2^{m-3+t}$ & $2^{m-t}-(n+1)\cdot 2^{1-m-t}-(2^{3-2m}\cdot(n+1)^2-4\cdot(n+1))\cdot2^{-2t}$ \\
     \hline
\end{tabular}
\end{center}}
\end{table}
\item{{\rm (2)}}   If $\epsilon=1$, then $\mathcal{C}_{D_{ 1}}$  is an $[n, 2m]$ code with weight distribution in Table \ref{Table2}, where
    $n=2^{2m-2}+\Phi_{1}.$ Its dual code has parameters $[n, n-2m, 4]$, which is distance-optimal with respect to the Sphere Packing bound.
    \begin{table}[h]
{\caption{\rm   The weight distribution of $\mathcal{C}_{D_{1}}$}\label{Table2}
\begin{center}
\begin{tabular}{cccc}\hline
     Weight & Multiplicity \\\hline
  $0$ & $1$  \\
 $\frac{n}{2}$ & $(2^{5-2m}\cdot n^2-2^4\cdot n)\cdot 2^{-2t}+2^{2m}-2$ \\
  $\frac{n}{2} + 2^{m-3+t}$ & $(2^3\cdot n-2^{4-2m}\cdot n^2)\cdot2^{-2t}$ \\
  $\frac{n}{2} - 2^{m-3+t}$ & $(2^3\cdot n-2^{4-2m}\cdot n^2)\cdot2^{-2t}$ \\
  $n$ & $1$\\
     \hline
\end{tabular}
\end{center}}
\end{table}
\end{description}
\end{theorem}
{\it Proof.}
 We only prove the weight distribution of $\mathcal{C}_{D_{\epsilon}}$ for the case~(i).  The weight distribution of $\mathcal{C}_{D_{\epsilon}}$ can be shown similarly if the condition (ii) holds.
The proof will be divided into two cases.

\noindent{\bf Case 1:} $\epsilon=0$. From (\ref{eq:wt0}) and (\ref{eq:wt01}), we obtain $\wt(\mathbf{c}(a,b))=0$ if $(a,b)=(0,0)$ or $(a,b)=(1,1)$. This means that every codeword
in $\mathcal{C}_{D_{ 0}}$ at least repeats $2$ times, i.e., $\mathcal{C}_{D_{ 0}}$ is degenerate and its dimension is less than or equal to $2m-1$.
From  (\ref{eq:wt0}) and (\ref{eqNab}),  the dimension of  $\mathcal{C}_{D_{ 0}}$ is less than $2m-1$ if and only if there exists a pair $(a,b) \in (\mathbb{F}_{2^m}, \mathbb{F}_{2^m})\backslash \{(0,0),(1,1)\}$ such that
\begin{equation}\label{n0ddd}
\begin{split}
n+1=\frac{1}{4}\left(\hat{f}(a)\hat{g}(b)+\hat{f}(a+1)\hat{g}(b+1)\right).
\end{split}
\end{equation}
On the other hand, we know
$$\frac{1}{4}\left(\hat{f}(a)\hat{g}(b)+\hat{f}(a+1)\hat{g}(b+1)\right) \in \left\{0, \pm2^{m+\frac{d_0+d_1}{2}-2}\right\}$$
since $\hat{f}(a)\cdot \hat{f}(a+1)=0$, $\hat{f}(a) \in \{0, \pm2^{\frac{m+d_0}{2}}\}$  and $\hat{g}(b)\in \{0, \pm2^{\frac{m+d_1}{2}}\}$ for any $a,b \in \mathbb{F}_{2^m}$. As $n$ is the length of $\mathcal{C}_{D_{0}}$, then  (\ref{n0ddd}) holds if and only if
$n+1=2^{m+\frac{d_0+d_1}{2}-2},$
which is impossible since $n=2^{2m-2}+\Phi_{0}-1$ and $\Phi_{0}\neq 2^{m+\frac{d_0+d_1}{2}-2}-2^{2m-2}$. Hence, in this case, the dimension of $\mathcal{C}_{D_{ 0}}$ is $2m-1$.
In the following, we determine the weight distribution of $\mathcal{C}_{D_{0}}$.

As $\hat{f}(a)\hat{g}(b)+\hat{f}(a+1)\hat{g}(b+1) \in \{0, \pm2^{m+\frac{d_0+d_1}{2}}\}$
for $(a,b)\in \bF_{2^m}^2\setminus\{ (0,0), (1,1)\}$, then from (\ref{lengthn}) and Proposition \ref{pro1}, the possible weights of $\mathcal{C}_{D_{ 0}}$ are
$$\left\{0, \,\,\frac{n+1}{2},\,\, \frac{n+1}{2} \pm 2^{m+\frac{d_0+d_1}{2}-3}\right\}.$$
Assume that $w_0=\frac{n+1}{2}$, $w_1=\frac{n+1}{2} + 2^{m+\frac{d_0+d_1}{2}-3}$ and $w_2=\frac{n+1}{2} - 2^{m+\frac{d_0+d_1}{2}-3}$. Let $A_{w_i}$ be the number of the codewords with weight $w_i$ in $\mathcal{C}_{D_{ 0}}$, where $0\leq i \leq 2$. It is clear that the dual code of $\mathcal{C}_{D_{ 0}}$
has the minimum weight at least $3$, from the first three Pless power moments identities, we have
\[\left\{ \begin{array}{lll}
\sum_{i=0}^2{\omega_i}=2^{2m-1}-1,\\
\sum_{i=0}^2\omega_iA_{\omega_i}=2^{2m-2}n,\\
\sum_{i=0}^2\omega_i^2A_{\omega_i}=2^{2m-3}n(n+1).\\
\end{array}\right. \]
Solving this system of equations, we obtain
\[\left\{ \begin{array}{lll}
A_{\omega_0}=2^{4-d_0-d_1-2m}(n+1)^2+2^{2m-1}-2^{3-d_0-d_1}\cdot(n+1)-1,\\
A_{\omega_1}=(n+1)\cdot 2^{1-m-\frac{d_1+d_2}{2}}-2^{m-\frac{d_1+d_2}{2}}-(2^{3-2m}\cdot(n+1)^2-4\cdot(n+1))\cdot2^{-d_0-d_1},\\
A_{\omega_2}=2^{m-\frac{d_1+d_2}{2}}-(n+1)\cdot 2^{1-m-\frac{d_1+d_2}{2}}-(2^{3-2m}\cdot(n+1)^2-4\cdot(n+1))\cdot2^{-d_0-d_1}.\\
\end{array}\right. \]

From the fourth Pless power moments identities, we have the number of the codewords of $\mathcal{C}_{D_{ 0}}^{\perp}$ with Hamming weight $3$ is
\begin{equation}\label{eq:B3}
B_3=\frac{2^{2m+d_0+d_1-4}+(n+1)^3\cdot 2^{1-2m}-(n+1)\cdot 2^{d_0+d_1-3}-3n-1}{6}.
\end{equation}
By the definition of $n$, it is easy to see
\begin{equation*}
n=\left\{ \begin{array}{lll}
2^{2m-2}-2^{m+\frac{d_0+d_1}{2}-2}-1, \,\,&\text{ if $\Phi_{0}=-2^{m+\frac{d_0+d_1}{2}-2}$},\\
2^{2m-2}-1, &\text{ if $\Phi_{0}=0$},\\
2^{2m-2}+2^{m+\frac{d_0+d_1}{2}-2}-1, \,\,&\text{ if $\Phi_{0}=2^{m+\frac{d_0+d_1}{2}-2}$}.\\
\end{array}\right.
\end{equation*}
Substituting the value of $n$ into (\ref{eq:B3}), we can check that $B_3\neq 0$ for $m\geq 3$.  This means that $d_H(\mathcal{C}_{D_{ 0}}^{\perp})= 3$.

\noindent{\bf Case 2:} $\epsilon=1$.  From (\ref{eq:wt0}) and (\ref{eq:wt01}), we obtain $\wt(\mathbf{c}(a,b))=0$ for $(a,b)=(0,0)$ and  $\wt(\mathbf{c}(a,b))=n$ for $(a,b)=(1,1)$.  By a similar argument as in Case~1, we see that for any $(a,b) \in \mathbb{F}_{2^m}^2\backslash \{(0,0),(1,1)\}$, the possible values of $\wt(\mathbf{c}(a,b))$ are
$$\left\{ \frac{n}{2},\,\, \frac{n}{2} \pm 2^{m+\frac{d_0+d_1}{2}-3}\right\},$$
which all are nonzero. This means that the dimension of $\mathcal{C}_{D_{ 1}}$ is $2m$.

Assume that $w_0=\frac{n}{2}$, $w_1=\frac{n}{2} + 2^{m+\frac{d_0+d_1}{2}-3}$ and $w_2=\frac{n}{2} - 2^{m+\frac{d_0+d_1}{2}-3}$. We now determine the number $A_{w_i}$ of codewords with weight $w_i$ in $\mathcal{C}_{D_{ 1}}$, where $0\leq i \leq 2$. It is clear that the dual code $\mathcal{C}_{D_{1}}^{\perp}$ of $\mathcal{C}_{D_{ 1}}$
has the minimum distance at least $3$, then the first three Pless power moments identities lead to the following system of equations:
\[\left\{ \begin{array}{lll}
\sum_{i=0}^2A_{\omega_i}=2^{2m}-2,\\
\sum_{i=0}^2\omega_iA_{\omega_i}+n=2^{2m-1}n,\\
\sum_{i=0}^2\omega_i^2A_{\omega_i}+n^2=2^{2m-2}n(n+1).\\
\end{array}\right. \]
Solving this system of equations, we obtain
\[\left\{ \begin{array}{lll}
A_{\omega_0}=(2^{5-2m}\cdot n^2-2^4\cdot n)\cdot 2^{-d_0-d_1}+2^{2m}-2,\\
A_{\omega_1}=(2^3\cdot n-2^{4-2m}\cdot n^2)\cdot2^{-d_0-d_1},\\
A_{\omega_2}=(2^3\cdot n-2^{4-2m}\cdot n^2)\cdot2^{-d_0-d_1}.\\
\end{array}\right. \]

Next, we show that the minimum distance of $\mathcal{C}_{D_{ 1}}^{\perp}$ is $4$. 
Assume that $\mathcal{C}_{D_{ 1}}^{\perp}$ has a codeword $\mathbf{c}$ with Hamming weight $3$. By Proposition \ref{pro1}, we know that $(1,1, \dots, 1)$ is a codeword in $\mathcal{C}_{D_{ 1}}$ since which is an only codeword with weight $n$. So, $\mathbf{c} \cdot (1,1,\cdots, 1)=0$. This is a contradiction.
 Hence, $d_H(\mathcal{C}_{D_{ 1}}^{\perp})\geq 4$. If $d_H(\mathcal{C}_{D_{ 1}}^{\perp})=5$, from Sphere Packing bound, we have
\begin{equation}\label{2mm}
2^{n} \geq 2^{n-2m}\sum_{i=0}^2\left( \begin{array}{cccc}
   n  \\
     i  \\
\end{array}
\right), \,\, i.e.,   \,\,   2^{2m} \geq 1+n+\frac{n(n-1)}{2}.
\end{equation}
It is easy to check that (\ref{2mm}) does not hold for $m\geq 3$. This means that $d_H(\mathcal{C}_{D_{ 1}}^{\perp})=4$. So, $\mathcal{C}_{D_{ 1}}^{\perp}$ has parameters $\left[n,n-2m,4\right]$ and is distance-optimal with respect to the Sphere Packing bound. $\square$

\begin{remark}
Note that almost all known Boolean functions $f(x)$ and $g(x)$ with the condition (i) or (ii) in Theorem \ref{theorem1} satisfy $\Phi_{\epsilon}\neq 2^{m+t-2}-2^{2m-2}$,
i.e., the dimensions of the codes $\mathcal{C}_{D_0}$ and $\mathcal{C}_{D_1}$ are $2m-1$ and $2m$ respectively, where $\Phi_{\epsilon}$ is defined in (\ref{eq:phi1}).
On the other hand, from Table~\ref{Table1} and Table~\ref{Table2}, we see that the Hamming weights of all codeword in $\mathcal{C}_{D_{ \epsilon}}$ are related to the Walsh
transform values of $f(x)$ and $g(y)$. These values can be obtained explicitly for some Boolean functions $f(x)$ and $g(y)$ in the following corollaries.
\end{remark}

\begin{corollary}\label{cor1}
Let $m, k$ be positive integers with $m\equiv 2 \pmod 4$ and $d=\gcd(m,k)$ being odd. Let $\mathcal{C}_{D_{ \epsilon}}$ be a linear code with the defining set $D_{ \epsilon}$ given in~(\ref{eq:defc1}), where $f(x)={\rm Tr}_1^m(x^{2^k+1})$ and $g(y)={\rm Tr}_1^m(y^e)$ with $e=2^{\frac{m}{2}}+2^{\frac{m+2}{4}}+1$ or $e=2^{\frac{m+2}{2}}+3$.
Then the following statements hold.
\begin{description}
\item{\rm (1)} If $\epsilon=0$, then $\mathcal{C}_{D_{0}}$ is a $[2^{2m-2}-1,2m-1,2^{2m-3}-2^{m+d-2}]$ code with weight enumerator
\[ 1+(2^{2m-2d-3}+2^{m-d-2})x^{2^{2m-3}-2^{m+d-2}}+(2^{2m-1}-2^{2m-2d-2}-1)x^{2^{2m-3}}+(2^{2m-2d-3}-2^{m-d-2})x^{2^{2m-3}+2^{m+d-2}}. \]

\item{\rm (2)}  If $\epsilon=1$, then $\mathcal{C}_{D_{1}}$  is a $[2^{2m-2},2m,2^{2m-3}-2^{m+d-2}]$ code with weight enumerator
\[ 1+(2^{2m}-2^{2m-2d+1}-2)x^{2^{2m-3}}+2^{2m-2d-2}\left(x^{2^{2m-3}-2^{m+d-2}}+x^{2^{2m-3}+2^{m+d-2}}\right)+x^{2^{2m-2}}. \]
\end{description}
\end{corollary}
{\it Proof.} From \cite{Cusick1996}, we know that $g(y)={\rm Tr}_1^m(y^e)$ is a plateaued function for $e=2^{\frac{m}{2}}+2^{\frac{m+2}{4}}+1$ or $e=2^{\frac{m+2}{2}}+3$, and $\hat{g}(\omega)\in \{0, \pm 2^{\frac{m+2}{2}}\}$ for any $\omega \in \mathbb{F}_{2^m}$.
It is easy to verify that $\gcd( e, 2^m-1)=1$, and so $\hat{g}(0)=0$. On the other hand, from Lemma \ref{beglem12} we know that $\hat{f}(1)=0$ and $\hat{f}(\omega) \in \{0, \pm 2^{\frac{m+2d}{2}}\}$ for any $\omega \in \mathbb{F}_{2^m}$. By Theorem~\ref{theorem1}, the length of the code $n=2^{2m-2}-1$ or $2^{2m-2}$ if $\epsilon=0$ or $1$, respectively. Moreover, $f(x)$ and $g(y)$ satisfy the condition (i) in Theorem \ref{theorem1}. Substituting the values of $n$ and $t=d+1$ into Table~\ref{Table1} and Table~\ref{Table2}, we get the weight enumerators in (1) and (2), respectively. $\square$

\begin{corollary}\label{cor2}
Let $m, k$ be positive integers with $m\equiv 2 \pmod 4$ and  $d=\gcd(m,k)$ being odd. Let $\mathcal{C}_{D_{ \epsilon}}$ be a linear code with the defining set $D_{ \epsilon}$ given by~(\ref{eq:defc1}) in which $f(x)={\rm Tr}_1^m(x^{2^k+1})$ and $g(y)={\rm Tr}_1^m(\alpha y^e)$, where $\alpha$ and $e$ satisfying one of the following conditions:
\begin{description}
\item{$\bullet$} $e=2^{h}+1$, where $h$ is a positive integer and $\alpha \notin \{ x^e\, | \, x\in \mathbb{F}_{2^m}\}$;
\item{$\bullet$} $e=2^{2h}-2^h+1$, where $\gcd(h,m)=1$ and $\alpha \notin \{ x^3\, | \, x\in \mathbb{F}_{2^m}\}$;
\item{$\bullet$} $e=2^{h}-1$, where $h\geq 2$ and $\alpha$ is a zero of the Kloosterman Sum.
\end{description}
Denote $\mu=1$ if $\gcd(e, 2^{\frac{m}{2}}-1)=1$ and $\mu=-1$ if $\gcd(e, 2^{\frac{m}{2}}+1)=1$, then the following statements hold.
\begin{description}
 \item{\rm (1)} If $\epsilon=0$, then $\mathcal{C}_{D_{0}}$ is a $[2^{2m-2}-\mu2^{m+d-2}-1,2m-1,2^{2m-3}-(1+\mu)2^{m+d-3}]$ code with weight enumerator
   $$ 1+(2^{2m-2d-1}-\mu2^{m-d-1}-1)x^{2^{2m-3}}+(2^{2m-1}-2^{2m-2d})x^{2^{2m-3}-\mu2^{m+d-3}}+(2^{2m-2d-1}+\mu2^{m-d-1})x^{2^{2m-3}-\mu2^{m+d-2}}.$$
\item{\rm (2)} If $\epsilon=1$, then $\mathcal{C}_{D_{1}}$ is a $[2^{2m-2}-\mu2^{m+d-2},2m,2^{2m-3}-(1+\mu)2^{m+d-3}]$ code with weight enumerator
\[ 1+(2^{2m-2d}-1)\left(x^{2m-3}+x^{2^{2m-3}-\mu2^{m+d-2}}\right)+(2^{2m}-2^{2m-2d+1})x^{2^{2m-3}-\mu2^{m+d-3}}+x^{2^{2m-2}-\mu2^{m+d-2}}. \]
\end{description}
\end{corollary}
{\it Proof.} It is easy to see that $\gcd(e, 2^{\frac{m}{2}}-1)=1$ or $\gcd(e, 2^{\frac{m}{2}}+1)=1$ since $\gcd\left(2^{\frac{m}{2}}-1, 2^{\frac{m}{2}}+1\right)=1$. In the following, we only consider the case $\gcd(e, 2^{\frac{m}{2}}-1)=1$ and the other case can be shown similarly.

From \cite{ Dillon1974, DD2004}, we know that $g(y)={\rm Tr}_1^m(\alpha y^e)$ is a bent function for all $e$ listed above, and so $\hat{g}(\omega) \in \{ \pm 2^{\frac{m}{2}}\}$.
Since $\gcd\left(e, 2^{\frac{m}{2}}-1\right)=1$, we have $s =\gcd(e, 2^{\frac{m}{2}}+1)\neq 1 $. Let $\gamma$ be a primitive element of $\bF_{2^m}$ and $G=\langle\gamma^s\rangle$
be a subgroup of $\bF_{2^m}^*$ with order $(2^m-1)/s$. Then
\begin{equation*}\label{eqfff}
\hat{g}(0)= \sum_{y \in \mathbb{F}_{2^m}}(-1)^{{\rm Tr}_1^m\left(y^e\right)}=1+\sum_{y \in \mathbb{F}_{2^m}^*}(-1)^{{\rm Tr}_1^m\left(y^e\right)}= 1+s\sum_{y \in G}(-1)^{{\rm Tr}_1^m\left(y^{e}\right)}\equiv 1 \pmod {s}.
\end{equation*}
So, $\hat{g}(0)=-2^{\frac{m}{2}}$. On the other hand, from Theorem \cite[Theorem 5.2]{Coulter1999}, we have $\hat{f}(0)=2^{\frac{m+2d}{2}}$ and from Lemma~\ref{beglem12}, we obtain that $\hat{f}(1)=0$ and $\hat{f}(\omega) \in \{0, \pm 2^{\frac{m+2d}{2}}\}$ for any $\omega \in \mathbb{F}_{2^m}$. By Theorem~\ref{theorem1}, the length of the code
$n=2^{2m-2}-2^{m+d-2}-1$ or $2^{2m-2}-2^{m+d-2}$ if $\epsilon=0$ or $1$, respectively. It is obvious that $f(x)$ and $g(y)$ satisfy the condition (i) in Theorem \ref{theorem1}. Substituting the values of $n$ and $t=d$ into Table \ref{Table1} and Table \ref{Table2}, we obtain the weight enumerators in (1) and (2), respectively. $\square$

\begin{corollary}\label{cor3}
Let $m$ be an odd number with $m \geq 3$ and $3 \nmid m$. Let $k$ be a positive integer with $\gcd(m,k)=1$ and $\mathcal{C}_{D_{ \epsilon}}$ be a linear code with the defining set $D_{ \epsilon}$ given by~(\ref{eq:defc1}) in which $f(x)={\rm Tr}_1^m(x^{2^k+1}+x^{2^{2k}+1})$ and $g(y)={\rm Tr}_1^m(y^e)$, where $e$ is one of the following number:
\begin{description}
\item{$\bullet$} $e=2^{\frac{m-1}{2}}+3$, or $e=2^{2h}-2^h+1$, or $e=2^h+1$ for $\gcd(m,h)=1$;
\item{$\bullet$} $e=2^{\frac{m-1}{2}}+2^{\frac{m-1}{4}}-1$ for $m\equiv 1 \pmod 4$, or $e=2^{\frac{m-1}{2}}+2^{\frac{3m-1}{4}}-1$ for $m\equiv 3 \pmod 4$.
\end{description}
Then the following statements hold.
\begin{description}
\item{\rm (1)} If $\epsilon=0$, then $\mathcal{C}_{D_{0}}$ is a $[2^{2m-2}-1, \,\,2m-1,\,\, 2^{2m-3}-2^{m-2}]$ code with weight enumerator
    \[ 1+(2^{2m-3}+2^{m-2})x^{2^{2m-3}-2^{m-2}}+(2^{2m-2}-1)x^{2^{2m-3}}+(2^{2m-3}-2^{m-2})x^{2^{2m-3}+2^{m-2}}.\]
\item{\rm (2)} If $\epsilon=1$, then $\mathcal{C}_{D_{1}}$  is a $[2^{2m-2}, \, \,2m,\,\, 2^{2m-3}-2^{m-2}]$ code with weight enumerator
\[ 1+2^{2m-2}x^{2^{2m-3}-2^{m-2}}+(2^{2m-1}-2)x^{2^{2m-3}}+2^{2m-2}x^{2^{2m-3}+2^{m-2}}+x^{2^{2m-2}}. \]
\end{description}
\end{corollary}
{\it Proof.}
It is easy to verify that $\gcd(e, 2^m-1)=1$, and so $\hat{g}(0)=0$. From \cite{Gold1968,Kasami1971,Hollmann2001} we know that $g(y)={\rm Tr}_1^m(y^e)$ is a semi-bent function for all $e$ listed above, and so $\hat{g}(\omega)\in \{0, \pm 2^{\frac{m+1}{2}}\}$ for any $\omega \in \mathbb{F}_{2^m}$. From Lemma \ref{begle} we know that $\hat{f}(1)=0$.
By Theorem~\ref{theorem1}, the length of the code $n=2^{2m-2}-1$ or $2^{2m-2}$ if $\epsilon=0$ or $1$, respectively. It is easy to see that $f(x)$ and $g(y)$ satisfy the condition (i) in Theorem \ref{theorem1}. Substituting the values of $n$ and $t=1$ into Table \ref{Table1} and Table \ref{Table2}, we obtain the weight enumerators in ${\rm (1)}$
and ${\rm (2)}$, respectively. $\square$

\begin{corollary}\label{cor4}
Let $m$ be an integer with $m\geq 3$ and $\mathcal{C}_{D_{ \epsilon}}$ be a linear code with the defining set $D_{ \epsilon}$ given in~(\ref{eq:defc1}). If
$f(x)$ and $g(y)$ in~(\ref{eq:defc1}) are the same bent functions, then the following statements hold.
\begin{description}
\item{\rm (1)} If $\epsilon=0$, then $\mathcal{C}_{D_{0}}$ is a $[2^{2m-2}+2^{m-1}-1,\,\, 2m-1,\,\, 2^{2m-3}]$ code with weight enumerator
    $$ 1+(2^{2m-3}+2^{m-2}-1)x^{2^{2m-3}}+2^{2m-2}x^{2^{2m-3}+2^{m-2}}+(2^{2m-3}-2^{m-2})x^{2^{2m-3}+2^{m-1}}.$$
\item{\rm (2)}  If $\epsilon=1$, then $\mathcal{C}_{D_{1}}$  is a $[2^{2m-2},\,\, 2m,\,\, 2^{2m-3}-2^{m-2}]$  code with weight enumerator
 $$1+(2^{2m-1}-2)x^{2^{2m-3}}+2^{2m-2}x^{2^{2m-3}-2^{m-2}}+2^{2m-2}x^{2^{2m-3}+2^{m-2}}+x^{2^{2m-2}}.$$
\end{description}
\end{corollary}
{\it Proof.} As $f(x)$ and $g(y)$ are the same bent functions, they satisfy the condition (ii) in Theorem \ref{theorem1}. So, the length of the code
 $n=2^{2m-2}+2^{m-1}-1$ or $2^{2m-2}$ if $\epsilon=0$ or $1$, respectively. Substituting the values of $n$ and $t=1$ into Table \ref{Table1} and Table \ref{Table2},
 we obtain the weight enumerators in $(1)$ and $(2)$, respectively.  $\square$

The following numerical examples show that many best codes can be obtained from our constructions.

\begin{example}\label{example1}
Let $\mathcal{C}_{D_{ \epsilon}}$ be a linear code with the defining set $D_{ \epsilon}$ given in~(\ref{eq:defc1}), where $f(x)={\rm Tr}_1^3(x^3)$ and $g(y)={\rm Tr}_1^3(y^3)$ are Boolean functions from $\bF_{2^3}$ to $\bF_2$. By Lemma \ref{beglem12} and Theorem \ref{theorem1}, then the following results hold.
\begin{description}
\item{\rm (1)} The linear code $\mathcal{C}_{D_{0}}$ has parameters $[19,5,8]$ and its dual has parameters $[19,14,3]$.
\item{\rm (2)} The linear code $\mathcal{C}_{D_{1}}$ has parameters $[20,6,8]$  and its dual has parameters $[20,14,4]$.
\end{description}
These codes and their duals are optimal respect to the tables of best codes known maintained at http://www.codeta-bles.de. These results are verified by Magma programs.
\end{example}

\begin{example}\label{example2}
Let $\mathcal{C}_{D_{ \epsilon}}$ be a linear code with the defining set $D_{ \epsilon}$ given in~(\ref{eq:defc1}), where $f(x)={\rm Tr}_1^4(\alpha x^3)$ and $g(y)={\rm Tr}_1^4(\alpha y^3)$ with $\alpha$ being a primitive element of $\mathbb{F}_{2^4}$. Then $f(x)$ and $g(y)$ are bent functions. By Theorem~\ref{theorem1} and Corollary~\ref{cor4}, the following results hold.
\begin{description}
\item{\rm (1)} The linear code $\mathcal{C}_{D_{0}}$ has parameters $[71,7,32]$ and its dual has parameters $[71,64,3]$.
\item{\rm (2)} The linear code $\mathcal{C}_{D_{1}}$ has parameters $[64,8,28]$ and its dual has parameters $[64,56,4]$.
\end{description}
These codes and their duals are optimal or almost optimal respect to the tables of best codes known maintained at http://www.codetables.de.
These results are verified by Magma programs.
\end{example}

\section{The weight distribution of the second class of linear codes}

In this section, we investigate the weight distribution of the linear code $\mathcal{C}_{D_{ \epsilon}}$, where $\mathcal{C}_{D_{ \epsilon}}$ has the form (\ref{eq:defc}) and $D_{ \epsilon}$ is defined in (\ref{eq:defc2}).
Assume that $n=|D_{ \epsilon}|$ is the length of $\C_{D_{ \epsilon}}$,
then
$$n=\sum_{(x,y)\in \mathbb{F}_{2^m}^2\setminus\{(0,0)\}} \left({\frac{1}{2}}\sum_{z_0 \in \mathbb{F}_{2}}(-1)^{z_0(f(x)+g(y))}\right)\left({\frac{1}{2}}\sum_{z_1 \in \mathbb{F}_{2}}(-1)^{z_1({\rm Tr}_1^m(x))}\right)\left({\frac{1}{2}}\sum_{z_2 \in \mathbb{F}_{2}}(-1)^{z_2({\rm Tr}_1^m(y)-\epsilon)}\right).$$
By a similar argument as in (\ref{lengthn}) we get
\begin{equation}\label{lengthng1}
n=2^{2m-3}+\frac{1}{8}\hat{f}(0)\hat{g}(0)+\frac{1}{8}\hat{f}(1)\hat{g}(0)+\frac{1}{8}(-1)^{\epsilon}\hat{f}(0)\hat{g}(1)+\frac{1}{8}(-1)^{\epsilon}\hat{f}(1)\hat{g}(1)-\delta,
\end{equation}
where $\delta$ is given in (\ref{eq:11t}). For any  $(a,b)\in \mathbb{F}_{2^m}^2$, the Hamming weight of a codeword $\mathbf{c}(a,b)=\left( {\rm Tr}_1^m(ax+by)\right)_{(x,y)\in D_{ \epsilon}}$ in $\mathcal{C}_{D_{ \epsilon}}$ is
\begin{equation}\label{eq:wt011}
{\rm wt_H}( {\bf c}(a,b))=n-N(a,b),
\end{equation}
where $n$ is the length of $\mathcal{C}_{D_{\epsilon}}$ and
\begin{equation*}
\begin{split}
N(a,b)&=\left|\left\{(x,y)\in \mathbb{F}_{2^m}^2\setminus\{(0,0)\}:\,\, f(x)+f(y)=0, \,\,{\rm Tr}_1^m(x)=0,\,\,{\rm Tr}_1^m(y)=\epsilon \,\, \text{and}\,\, {\rm Tr}_1^m\left(ax+by\right)=0 \right\}\right|. \\
\end{split}
\end{equation*}
From the definition of $n$, it is easy to see that
\begin{equation}\label{Nab01}
N(1,0)=n \,\, {\rm and } \,\, N(1,1)= N(0,1)=\left\{ \begin{array}{lcl}
n, \,\,\text{if $\epsilon=0$}, \vspace{2mm}\\
0, \,\,\text{if $\epsilon=1$.}
\end{array}\right.
\end{equation}
If $(a,b)\in \bF_{2^m}^2\setminus \{ (0,0), (1,0), (0,1), (1,1)\}$, then
\begin{equation*}
\begin{split}
&N(a,b)=\sum_{(x,y)\in \mathbb{F}_{2^m}^2\setminus\{(0,0)\}} \left({\frac{1}{2}}\sum_{z_0 \in \mathbb{F}_{2}}(-1)^{z_0(f(x)+f(y))}\right)\left({\frac{1}{2}}\sum_{z_1 \in \mathbb{F}_{2}}(-1)^{z_1({\rm Tr}_1^m(x))}\right)\\
&\hskip 3.5cm\left({\frac{1}{2}}\sum_{z_2 \in \mathbb{F}_{2}}(-1)^{z_2({\rm Tr}_1^m(y)-\epsilon)}\right)\left({\frac{1}{2}}\sum_{z_3 \in \mathbb{F}_{2}}(-1)^{z_3{\rm Tr}_1^m(ax+by)}\right).\\
\end{split}
\end{equation*}
 By a similar argument as in (\ref{eqNab}), we obtain
\begin{equation}\label{eqNab1g}
\begin{split}
N(a,b)
=\frac{n-\delta}{2}+\frac{1}{16}\hat{f}(a)\hat{f}(b)+\frac{1}{16}\hat{f}(a+1)\hat{f}(b)
+\frac{1}{16}(-1)^{\epsilon}\hat{f}(a)\hat{f}(b+1))
+\frac{1}{16}(-1)^{\epsilon}\hat{f}(a+1)\hat{f}(b+1),
\end{split}
\end{equation}
where $\delta$ is given in (\ref{eq:11t}).

With the above preparations, we have the following results.
\begin{proposition}\label{pro2}
Follow the notation introduced above.  Assume that $(a,b) \in \mathbb{F}_{2^m}^2\setminus\{(0,0),(0,1), (1,0), (1,1)\}$. If $\epsilon=0$,  then $\mathcal{C}_{D_{0}}$ is a binary linear code of length $n$ and its Hamming weights are given by the following multiset
\begin{equation*}
\left\{\frac{n+1}{2}-\frac{\hat{f}(a)\hat{g}(b)}{16}-\frac{\hat{f}(a+1)\hat{g}(b)}{16}
-\frac{\hat{f}(a)\hat{g}(b+1)}{16}-\frac{\hat{f}(a+1)\hat{g}(b+1)}{16}\,\right\}\bigcup\left\{0\right\}.
\end{equation*}
 If $\epsilon=1$, then $\mathcal{C}_{D_{1}}$ is a binary linear code of length $n$ and its Hamming weights are given by the following multiset
\begin{equation*}
\left\{\frac{n}{2}-\frac{\hat{f}(a)\hat{g}(b)}{16}-\frac{\hat{f}(a+1)\hat{g}(b)}{16}
+\frac{\hat{f}(a)\hat{g}(b+1)}{16}+
\frac{\hat{f}(a+1)\hat{g}(b+1)}{16} \right\}\bigcup\left\{0,n\right\}.
\end{equation*}
\end{proposition}

We now determine the weight distribution of $\mathcal{C}_{D_{ \epsilon}}$ from some special Boolean functions. For convenience, we write
\begin{equation*}
\Phi_{ \epsilon}=\frac{1}{8}\left(\hat{f}(0)\hat{g}(0)+\hat{f}(1)\hat{g}(0)+(-1)^{\epsilon}\hat{f}(0)\hat{g}(1) +(-1)^{\epsilon}\hat{f}(1)\hat{g}(1)\right), \,\, \epsilon \in \{ 0, 1\}.
\end{equation*}

\begin{theorem}\label{theorem2}
Let $m$ be an integer with $m\geq 3$ and $\mathcal{C}_{D_{ \epsilon}}$ be a linear code with the defining set $D_{ \epsilon}$ given in~(\ref{eq:defc2}). Let $f(x)$ and $g(y)$ in~(\ref{eq:defc2}) satisfy one of the following conditions:
 \begin{description}
\item{\rm (i)} $\hat{f}(a) \in \left\{0, \pm2^{\frac{m+d_0}{2}}\right\}$, $\hat{f}(a)\cdot \hat{f}(a+1)=0$, $\hat{g}(b)\in \left\{0, \pm2^{\frac{m+d_1}{2}}\right\}$ and $\hat{g}(b)\cdot \hat{g}(b+1)=0$ for any $a,b \in \mathbb{F}_{2^m}$;
\item{\rm (ii)}  $\hat{f}(a) \in \left\{0, \pm2^{\frac{m+d_0}{2}}\right\}$, $\hat{f}(a)\cdot \hat{f}(a+1)=0$,  $\hat{g}(b)\in \left\{0, \pm2^{\frac{m+d_1}{2}}\right\}$ and $\hat{g}(b)=\pm \hat{g}(b+1)$ for any $a,b \in \mathbb{F}_{2^m}$.
 \end{description}
Denote $t=\frac{d_0+d_1}{2}$ or $\frac{d_0+d_1+2}{2}$ if the condition {\rm (i)} or {\rm (ii)} holds, respectively. Assume that $\Phi_{ \epsilon}\neq 2^{m+t-3}-2^{2m-3}$, then the following statements holds.
\begin{description}
\item{\rm (1)} If $\epsilon=0$, then $\mathcal{C}_{D_{0}}$ is an $[n,\,\,2m-2]$ code with weight distribution in Table \ref{Table3}, where
  $n=2^{2m-3}+\Phi_{0}-1$. Its dual code has parameters $[n,\,\, n-2m+2,\,\, 3]$.
    \begin{table}[h]
{\caption{\rm   The weight distribution of $\mathcal{C}_{D_{0}}$ }\label{Table3}
\begin{center}
\begin{tabular}{cccc}\hline
     Weight & Multiplicity \\\hline
  $0$ & $1$  \\
 $\frac{n+1}{2}$ & $((n+1)^2 2^{6-2m}-2^4 n-2^4)\cdot 2^{-2t}+2^{2m-2}-1$ \\
  $\frac{n+1}{2} + 2^{m-4+t}$ & $(n+1)2^{2-m-t}+(1-2^{2-2m}(n+1)^2+n)\cdot 2^{3-2t}-2^{m-t}$ \\
    $\frac{n+1}{2} - 2^{m-4+t}$ & $2^{m-t}-(n+1)2^{2-m-t}+(1-2^{2-2m}(n+1)^2+n)\cdot 2^{3-2t}$  \\
     \hline
\end{tabular}
\end{center}}
\end{table}
\item{\rm (2)}  If $\epsilon=1$, then $\mathcal{C}_{D_{1}}$ is an $[n,\,\, 2m-1]$ code with weight distribution in Table \ref{Table4}, where
$n=2^{2m-3}+\Phi_{1}.$ Its dual code has parameters $[n,\,\, n-2m+1,\,\, 4]$, which is distance-optimal with respect to the Sphere Packing bound.
    \begin{table}[h]
{\caption{\rm   The weight distribution of $\mathcal{C}_{D_{1}}$}\label{Table4}
\begin{center}
\begin{tabular}{cccc}\hline
     Weight & Multiplicity \\\hline
  $0$ & $1$  \\
 $n/2$ & $(2^{7-2m}\cdot n^2-2^5\cdot n)\cdot 2^{-2t}+2^{2m-1}-2$ \\
  $n/2 + 2^{m-4+t}$ & $(2^4n-n^2\cdot 2^{6-2m})\cdot2^{-2t}$ \\
  $n/2 - 2^{m-4+t}$ & $(2^4n-n^2\cdot 2^{6-2m})\cdot2^{-2t}$ \\
  $n$ & $1$\\
     \hline
\end{tabular}
\end{center}}
\end{table}
\end{description}
\end{theorem}
{\it Proof.} We only prove the weight distribution of $\mathcal{C}_{D_{\epsilon}}$ for the case (i). The weight distribution of $\mathcal{C}_{D_{\epsilon}}$ in the case~(ii) can be derived similarly.
The proof falls into two cases.

\noindent {\bf Case 1:} $\epsilon=0$. From (\ref{eq:wt011}) and (\ref{Nab01}), we know that $\wt(\mathbf{c}(a,b))=0$ if $(a,b)\in \{ (0,0), (1,0), (0,1), (1,1)\}$. So, each codeword in $\mathcal{C}_{D_{0}}$ at least repeats $4$ times, i.e., $\mathcal{C}_{D_{0}}$ is degenerate and its dimension is less than or equal to $2m-2$.
From (\ref{eq:wt011}) and (\ref{eqNab1g}), we know that the dimension of  $\mathcal{C}_{D_{0}}$ is less than $2m-2$ if and only if there exists a pair $(a,b)\in \bF_{2^m}\setminus \{ (0,0), (1,0), (0,1), (1,1)\}$ such that
\begin{equation}\label{n1ddd}
n+1=\frac{1}{8}\left(\hat{f}(a)\hat{g}(b)+\hat{f}(a+1)\hat{g}(b)+\hat{f}(a)\hat{g}(b+1)+\hat{f}(a+1)\hat{g}(b+1)\right).
\end{equation}
As $\hat{f}(a)\cdot \hat{f}(a+1)=0$ and $\hat{g}(b)\cdot \hat{g}(b+1)=0$ for any $a, b\in \bF_{2^m}$, there is at most one nonzero term among
$\hat{f}(a)\hat{f}(b)$, $\hat{f}(a+1)\hat{f}(b)$, $\hat{f}(a)\hat{f}(b+1)$ and $\hat{f}(a+1)\hat{f}(b+1)$. So,
\begin{equation}\label{fdscdsx}
\hat{f}(a)\hat{f}(b)+\hat{f}(a+1)\hat{f}(b)+\hat{f}(a)\hat{f}(b+1)+\hat{f}(a+1)\hat{f}(b+1)\in \left\{0, \pm 2^{m+\frac{d_0+d_1}{2}}\right\}
\end{equation}
since $\hat{f}(a)\in \{0, \pm 2^{\frac{m+d_0}{2}}\}$ and $\hat{g}(b)\in \{0, \pm 2^{\frac{m+d_1}{2}}\}$  for any $a,b \in \mathbb{F}_{2^m}$.
This means that  (\ref{n1ddd}) holds if and only if
$n+1=2^{m+\frac{d_0+d_1}{2}-3},$
which is impossible since $n=2^{2m-3}+\Phi_{0}-1$ and $\Phi_{0}\neq 2^{m+\frac{d_0+d_1}{2}-3}-2^{2m-3}$.
Hence,  in this case, the dimension of $\mathcal{C}_{D_{0}}$ is $2m-2$.
In the following, we determine the weight distribution of $\mathcal{C}_{D_{0}}$.

By Proposition \ref{pro2} and (\ref{fdscdsx}), the set of possible Hamming weights of $\mathcal{C}_{D_{0}}$ is
$$\left\{0, \frac{n+1}{2}, \frac{n+1}{2} \pm 2^{m+\frac{d_0+d_1}{2}-4}\right\}.$$
Let $w_0=\frac{n+1}{2}$, $w_1=\frac{n+1}{2} + 2^{m+\frac{d_0+d_1}{2}-4}$ and $w_2=\frac{n+1}{2} - 2^{m+\frac{d_0+d_1}{2}-4}$. Assume that $A_{w_i}$ is the number of the codewords with weight $w_i$ in $\mathcal{C}_{D_{0}}$, where $0\leq i \leq 2$. It is easy to see that the minimum weight of the dual code of $\mathcal{C}_{D_{0}}$
is at least $3$. From the first three Pless power moments identities, we have
\[\left\{ \begin{array}{lll}
\sum_{i=0}^2A_{\omega_i}=2^{2m-2}-1,\\
\sum_{i=0}^2\omega_iA_{\omega_i}=2^{2m-3}n,\\
\sum_{i=0}^2\omega_i^2A_{\omega_i}=2^{2m-4}n(n+1).\\
\end{array}\right. \]
Solving this system of equations, we obtain
\[\left\{ \begin{array}{lll}
A_{\omega_0}=(2^{5-2m}\cdot n^2-2^4\cdot n)\cdot 2^{-d_0-d_1}+2^{2m-1}-2,\\
A_{\omega_1}=(n+1)2^{2-m-\frac{d_0+d_1}{2}}+(1-2^{2-2m}(n+1)^2+n)\cdot 2^{3-(d_0+d_1)}-2^{m-\frac{d_0+d_1}{2}},\\
A_{\omega_2}=2^{m-\frac{d_0+d_1}{2}}-(n+1)2^{2-m-\frac{d_0+d_1}{2}}+(1-2^{2-2m}(n+1)^2+n)\cdot 2^{3-(d_0+d_1)}.\\
\end{array}\right. \]
From the fourth Pless power moments, the number of the codewords of $\mathcal{C}_{D_{0}}^{\perp}$ with Hamming weight $3$ is
\begin{equation}\label{eq:B4}
B_3=\frac{(n+1)^32^{2-2m}-(2^2(n+1)-2^{2m})\cdot 2^{d_0+d_1-6}-3n-1}{6}.
\end{equation}
By the definition of $n$ and $(\ref{fdscdsx})$, we have
\begin{equation*}
n=\left\{ \begin{array}{lll}
2^{2m-3}-2^{m+\frac{d_0+d_1}{2}-3}-1, \,\,&\text{ if $\Phi_{0 }=-2^{m+\frac{d_0+d_1}{2}-3}$},\\
2^{2m-3}-1, &\text{ if $\Phi_{0 }=0$},\\
2^{2m-3}+2^{m+\frac{d_0+d_1}{2}-3}-1, \,\,&\text{ if $\Phi_{0 }=2^{m+\frac{d_0+d_1}{2}-3}$}.\\
\end{array}\right.
\end{equation*}
Substituting the values of $n$ into (\ref{eq:B4}), we can check that $B_3\neq 0$ for $m\geq 3$. Hence, $d_H(\mathcal{C}_{D_{0}}^{\perp})= 3$.

\noindent{\bf Case 2:} $\epsilon=1$.
By a similar analysis as in Case~1, we have that the dimension of $\mathcal{C}_{D_{1}}$ is $2m-1$ and the set of the possible nonzero weight in $\mathcal{C}_{D_{1}}$ is
$\{ \frac{n}{2}, \frac{n}{2} \pm 2^{m+\frac{d_0+d_1}{2}-4}, n\}$.

Assume that $w_0=\frac{n}{2}$, $w_1=\frac{n}{2} + 2^{m+\frac{d_0+d_1}{2}-4}$, $w_2=\frac{n}{2} - 2^{m+\frac{d_0+d_1}{2}-4}$ and $w_3=n$. Let $A_{w_i}$ denote the number of the codewords with weight $w_i$ in $\mathcal{C}_{D_{1}}$, where $0\leq i \leq 3$. It is clear that $A_{w_3}=1$ and the dual code of $\mathcal{C}_{D_{1}}$ has the minimum weight at least $3$. From the first three Pless power moments identities, we have
\[\left\{ \begin{array}{lll}
\sum_{i=0}^2A_{\omega_i}=2^{2m-1}-2,\\
\sum_{i=0}^2\omega_iA_{\omega_i}+n=2^{2m-2}n,\\
\sum_{i=0}^2\omega_i^2A_{\omega_i}+n^2=2^{2m-3}n(n+1).\\
\end{array}\right. \]
Solving this system of equations, we obtain
\[\left\{ \begin{array}{lll}
A_{\omega_0}=(2^{7-2m}\cdot n^2-2^5\cdot n)\cdot 2^{-d_0-d_1}+2^{2m}-2,\\
A_{\omega_1}=(2^4n-n^2\cdot 2^{6-2m})\cdot2^{-d_0-d_1},\\
A_{\omega_2}=(2^4n-n^2\cdot 2^{6-2m})\cdot2^{-d_0-d_1}.\\
\end{array}\right. \]
From Proposition~\ref{pro2}, we know that $n$ is a Hamming weight of a codeword in $\mathcal{C}_{D_{1}}$. By a similar discussion as Case~2 in Theorem \ref{theorem1},
we have that $d_H(\mathcal{C}_{D_{1}}^{\perp})= 4$. It is easy to see that the code with parameters $[n,\,\,n-2m+1,\,\,4]$ is distance-optimal with respect to the Sphere Packing bound.   $\square$

In fact, some Hamming weights occur zero time in Table \ref{Table3} and Table \ref{Table4} for some special Boolean functions. We presents
two-weight or three-weight linear code $\mathcal{C}_{D_{ \epsilon}}$ from some special Boolean functions $f(x)$ and $g(y)$. The following Corollary~\ref{cor5}
 can be derived directly from Lemma \ref{begle} and Theorem \ref{theorem2}.

\begin{corollary}\label{cor5}
Let $m, k$ and $\ell$ be positive integers with $m\geq 3$ and $\gcd(k\ell,m)=\gcd(k(\ell+1),m)=1$. Let $\mathcal{C}_{D_{ \epsilon}}$ be the linear code with the defining set $D_{ \epsilon}$ given in~(\ref{eq:defc2}), where $f(x)={\rm Tr}_1^m(\sum_{i=1}^{\ell}x^{2^{ik}+1})$ and $g(y)={\rm Tr}_1^m(\sum_{i=1}^{\ell}y^{2^{ik}+1})$. Then the following statements hold.
\begin{description}
\item{\rm (1)} If $\epsilon=0$, then $\mathcal{C}_{D_{0}}$ is a $[2^{2m-3}+2^{m-2}-1,2m-2,2^{2m-4}]$ code with weight enumerator
    $$1+(2^{2m-3}+2^{m-2}-1)x^{2^{2m-4}}+(2^{2m-3}-2^{m-2})x^{2^{2m-4}+2^{m-2}}.$$
\item{\rm (2)} If $\epsilon=1$, then $\mathcal{C}_{D_{1}}$ is a $[2^{2m-3}+(-1)^\ell2^{m-2},2m-1,2^{2m-4}+(-1)^\ell2^{m-2}]$ code with weight enumerator
      $$1+(2^{2m-2}-1)x^{2^{2m-4}+(-1)^\ell2^{m-2}}+(2^{2m-2}-1)x^{2^{2m-4}}+x^{2^{2m-4}+(-1)^\ell2^{m-2}}.$$
\end{description}
\end{corollary}


\begin{corollary}\label{cor6}
Let $m, k$ be positive integer with $m\equiv 2 \pmod 4$ and $d=\gcd(m,k)$ being odd. Let $\mathcal{C}_{D_{ \epsilon}}$ be the linear code with the defining set $D_{ \epsilon}$ given by~(\ref{eq:defc2}) in which $f(x)={\rm Tr}_1^{m}(x^{2^k+1})$ and $g(y)={\rm Tr}_1^{m}(\alpha y^{2^{\frac{m}{2}+1}})$, where $\alpha=\frac{1}{\gamma^{2^m-1}+1}+\frac{1}{\gamma^{2^{\frac{3m}{2}}-2^{\frac{m}{2}}}+1}$ and $\gamma$ is a primitive element of $\mathbb{F}_{2^{2m}}$.
Then the following statements hold.
\begin{description}
\item{\rm (1)} If $\epsilon=0$, then $\mathcal{C}_{D_{0}}$ is a $[2^{2m-3}-1,\, \, 2m-2,\,\, 2^{2m-4}-2^{m+d-3}]$ code with weight enumerator
   $$ 1+(2^{2m-2}-2^{m-2d-2}-1)x^{2^{2m-4}}+(2^{2m-2d-3}-2^{m-d-2})x^{2^{2m-4}+2^{m+d-3}}+(2^{2m-2d+3}+2^{m-d-2})x^{2^{2m-4}-2^{m+d-3}}.$$
\item{\rm (2)} If $\epsilon=1$, then $\mathcal{C}_{D_{1}}$ is a $[2^{2m-3}-2^{m+d-2},\,\,2m-1,\,\,2^{2m-4}-2^{m+d-2}]$ code with weight enumerator
\[ 1+(2^{2m-2d-2}-1)\left(x^{2m-4}+x^{2^{2m-4}-2^{m+d-2}}\right)+(2^{2m}-2^{2m-2d-1})x^{2^{2m-3}-2^{m+d-3}}+x^{2^{2m-3}-2^{m+d-2}}.\]
\end{description}
\end{corollary}
{\it Proof}. It is easy to see that
\begin{equation*}
\alpha^{2^m}=\frac{1}{\gamma^{1-2^m}+1}+\frac{1}{\gamma^{2^{\frac{m}{2}}-2^{\frac{3m}{2}}}+1}=\frac{\gamma^{2^m-1}+
\gamma^{2^{\frac{3m}{2}}-2^{\frac{m}{2}}}}{(\gamma^{2^m-1}+1)(\gamma^{2^{\frac{3m}{2}}-2^{\frac{m}{2}}}+1)}=\alpha.
\end{equation*}
This means that $\alpha\in \mathbb{F}_{2^m}$. Similarly, we can prove that $\alpha^{2^\frac{m}{2}}+\alpha=1$. It is clear that
\begin{equation}\label{eqdddd}
\begin{split}
\hat{g}(0)\hat{g}(1)&=\sum_{z \in \mathbb{F}_{2^m}}(-1)^{{\rm Tr_1^m}\left(\alpha z^{2^{\frac{m}{2}}+1}\right)}\sum_{x \in \mathbb{F}_{2^m}}(-1)^{{\rm Tr}_1^m\left(\alpha x^{2^{\frac{m}{2}}+1}+x\right)}
=\sum_{x, y \in \mathbb{F}_{2^m}}(-1)^{{\rm Tr}_1^m\left(\alpha(x+y)^{2^{\frac{m}{2}}+1}+\alpha x^{2^{\frac{m}{2}}+1}+ x\right)}\\
&=\sum_{x, y \in \mathbb{F}_{2^m}}(-1)^{{\rm Tr}_1^m\left(\alpha y^{2^{\frac{m}{2}}+1}+\alpha x^{2^{\frac{m}{2}}}y+\alpha xy^{2^{\frac{m}{2}}}+ x\right)} \\
&=\sum_{y \in \mathbb{F}_{2^m}}(-1)^{{\rm Tr}_1^m\left(\alpha y^{2^{\frac{m}{2}}+1}\right)}\sum_{x \in  \mathbb{F}_{2^m}}(-1)^{{\rm Tr}_1^m\left(\left(\alpha y+\alpha ^{2^{\frac{m}{2}}}y+1\right)x^{2^{\frac{m}{2}}}\right)}\\
&=(-1)^{{\rm Tr}_1^m(\alpha)}2^m=-2^m.
\end{split}
\end{equation}
The last equality is derived from the fact that $m\equiv 2 \pmod 4$ and $\alpha^{2^\frac{m}{2}}+\alpha=1$. From the proof of Corollary~\ref{cor2}, it is easy for us to get $\hat{g}(0)=-2^{\frac{m}{2}}$. Then, we have $\hat{g}(1)=2^{\frac{m}{2}}$. By similar computations as in~(\ref{eqdddd}), we obtain that $\hat{g}^2(\omega)=2^{m}$ for any $\omega \in \mathbb{F}_{2^m}$. Hence, we know that $g(y)$ is a bent function.

On the other hand, from Theorem \cite[Theorem 5.2]{Coulter1999}, we have $\hat{f}(0)=2^{\frac{m+2d}{2}}$ and from Lemma \ref{beglem12}, we obtain that $\hat{f}(1)=0$ and $\hat{f}(\omega) \in \{0, \pm 2^{\frac{m+2d}{2}}\}$ for any $\omega \in \mathbb{F}_{2^m}$. Then $n=2^{2m-3}-1$ if $\epsilon=0$ and $n=2^{2m-3}-2^{m+d-2}$ if $\epsilon=1$, where $n$ is defined in (\ref{lengthng1}). It is obvious that $f(x)$ and $g(y)$ satisfy the condition (ii) in Theorem \ref{theorem2}. Substitute the value of $n$ and $t=d+1$ into Table \ref{Table3} and Table \ref{Table4}, the desired conclusion then follows. $\square$

In the following, we give two examples for $f(x)$ and $g(y)$ satisfying condition (i) or condition (ii) in Theorem~\ref{theorem2}, respectively.
 \begin{example}\label{example3}
Let $\mathcal{C}_{D_{ \epsilon}}$ be a binary linear code with the defining set $D_{ \epsilon}$ given in~(\ref{eq:defc2}), where $f(x)={\rm Tr}_1^5(\sum_{i=1}^{\ell}x^{2^{i}+1})$ and $g(y)={\rm Tr}_1^5(\sum_{i=1}^{\ell}y^{2^{i}+1})$ are Boolean functions from $\bF_{2^5}$ to $\bF_2$. By Corollary~\ref{cor5} and Theorem~\ref{theorem2}, the following  results hold.
\begin{description}
\item{(1)} Let $\epsilon=0$ and $\ell=2$, then $\mathcal{C}_{D_{0}}$ has parameters $[135,8,64]$ and its  dual has parameters $[135,127,3]$.
\item{(2)} Let $\epsilon=1$ and $\ell=2$, then $\mathcal{C}_{D_{1}}$ has parameters $[136,9,64]$ and its dual has parameters $[136,127,4]$.
\item{(3)} Let $\epsilon=0$ and $\ell=1$, then $\mathcal{C}_{D_{0}}$ has parameters $[120,9,56]$ and its  dual has parameters $[120,111,4]$.
 \end{description}
These codes and their duals are optimal or almost optimal respect to the tables of best codes known maintained at http://www.codetables.de.
These results are verified by Magma programs.
\end{example}

\begin{example}\label{example4}
Let $\gamma$ be a primitive element of $\mathbb{F}_{2^{12}}$ and $\alpha=\frac{1}{\gamma^{63}+1}+\frac{1}{\gamma^{504}+1}$. Let $\mathcal{C}_{D_{ \epsilon}}$ be a binary linear code with the defining set $D_{ \epsilon}$ given in~(\ref{eq:defc2}), where $f(x)={\rm Tr}_1^6(x^3)$ and $g(y)={\rm Tr}_1^6(\alpha y^{9})$ are Boolean functions from $\bF_{2^6}$ to $\bF_2$. By Corollary~\ref{cor6} and Theorem~\ref{theorem2}, the following results hold.
\begin{description}
\item{(1)} Let $\epsilon=0$, then $\mathcal{C}_{D_{0}}$ has parameters $[511,10,240]$ and its  dual has parameters $[511,501,3]$.
\item{(2)} Let $\epsilon=1$, then $\mathcal{C}_{D_{1}}$ has parameters $[480,11,224]$ and its dual has parameters $[480,469,4]$.
 \end{description}
\end{example}


\section{Concluding remarks }\label{sec:concluding}

In this paper we constructed many classes of binary linear codes with few weights from some Boolean functions with at most three Walsh transform values. In order to improve the rate of the objective linear codes, we gave more restrictions on the defining set. The linear codes constructed in this paper seem new since the Hamming weights occur in the obtained linear codes are new. Specifically, the main results are summarized as follows:

\begin{description}
\item{$\bullet$} We provided two general constructions of binary linear codes with few weights from Boolean functions with at most three Walsh transform values (see Theorem \ref{theorem1} and Theorem \ref{theorem2}).

\item{$\bullet$}   We presented the weight distribution of $\mathcal{C}_{D_{\epsilon}}$ explicitly for many special Boolean functions $f(x)$ and $g(y)$ (see Corollary \ref{cor1}, Corollary \ref{cor2}, Corollary \ref{cor3}, Corollary \ref{cor4}, Corollary \ref{cor5} and Corollary \ref{cor6}).

\item{$\bullet$}   According to Codetable, we obtained some optimal and almost optimal linear codes (see Example \ref{example1}, Example \ref{example2} and Example \ref{example3}).

\item{$\bullet$} A binary linear code is called {\it self-complementary} if it contains all-one vector. The code $\mathcal{C}_{D_{1}}$ is  self-complementary code and the dual $\mathcal{C}_{D_{1}}^{\perp}$ is distance-optimal with respect to Sphere Packing bound in Section~4 and Section~5.

\end{description}

A linear code $\mathcal{C}$ is said to be {\it projective} if any two of its coordinates are linearly independent, or in other words, if the minimum
distance of $\mathcal{C}^{\perp}$ is at least $3$. Binary projective linear codes are very interesting due to their applications in many areas. All linear codes constructed in this paper are projective codes and may be used to construct association schemes~\cite{CA1984} and strongly regular graphs~\cite{CA1986}. Moreover, projective two-weight codes given in Corollary~\ref{cor5} may be related to other combinatorial objects, such as caps in projective spaces and combinatorial designs \cite{B2006}.

Some binary linear codes obtained in this paper can be used to construct secret sharing schemes with interesting access structures. Let $w_{min}$ and $w_{max}$ denote the minimum and maximum nonzero weights of a linear code $\mathcal{C}$, respectively. Ding and Ding~\cite{DDing2015} showed that if the linear code $\C$ with $w_{min}/w_{max}>\frac{1}{2}$, then the secret sharing scheme based on the dual code $\mathcal{C}^{\perp}$ has the nice access structure. When $\epsilon=0$, the linear codes constructed in Theorem \ref{theorem1} and Theorem \ref{theorem2} satisfy $w_{min}/w_{max}>\frac{1}{2}$ if $m>t+2$. It then follows that the dual codes of $\C_{D_0}$ in Theorem~\ref{theorem1} and Theorem~\ref{theorem2}
can be employed to obtain secret sharing schemes with interesting access structures.

\end{document}